\documentclass[aps,prb,twocolumn,showpacs,groupedaddress, amsmath]{revtex4}

\usepackage{graphicx}
\usepackage{amssymb}

\begin{document}


\title{Aging dynamics across a dynamic crossover line in three
dimensional short-range Ising spin glass
Cu$_{0.5}$Co$_{0.5}$Cl$_{2}$-FeCl$_{3}$ graphite bi-intercalation compound}


\author{Itsuko S. Suzuki }
\email[]{itsuko@binghamton.edu}
\affiliation{Department of Physics, State University of New York at 
Binghamton, Binghamton, New York 13902-6000}

\author{Masatsugu Suzuki }
\email[]{suzuki@binghamton.edu}
\affiliation{Department of Physics, State University of New York at 
Binghamton, Binghamton, New York 13902-6000}


\date{\today}

\begin{abstract}
Cu$_{0.5}$Co$_{0.5}$Cl$_{2}$-FeCl$_{3}$ graphite intercalation compound is
a three-dimensional short-range Ising spin glass with a spin freezing
temperature $T_{c}$ ($= 3.92 \pm 0.11$ K).  The stability of the spin glass
phase in the presence of a magnetic field $H$ is examined from (i) the spin
freezing temperature $T_{f}(\omega,H)$ at which the differential
field-cooled (FC) susceptibility $\partial M_{FC}(T,H)/\partial H$
coincides with the dispersion $\chi^{\prime}(\omega,T,H=0)$ with the
angular frequency $\omega$, and (ii) the time dependence of zero-field
cooled (ZFC) susceptibility $\chi_{ZFC}$ after a ZFC aging protocol with a
wait time $t_{W}$ ($= 1.0 \times 10^{4}$).  The relaxation rate $S(t)$ (=
d$\chi_{ZFC}$/d$\ln t$) exhibits a local maximum at a characteristic time
$t_{cr}$, reflecting non-equilibrium aging dynamics.  The peak time
$t_{cr}(T,H)$ decreases with increasing $H$ at the fixed temperature $T$
(2.9 K $\leq T<T_{c}$).  The spin freezing temperature $T_{f}(\omega,H)$
provides evidence for the instability of the spin glass phase in thermal
equilibrium in a finite magnetic field.  Both $t_{cr}(T,H)$ and
$T_{f}(\omega,H)$ exhibit certain scaling behavior predicted from the
droplet picture, suggesting a dynamic crossover from SG dynamics to
paramagnetic behavior in the presence of $H$.
\end{abstract}


\maketitle



\section{\label{intro}Introduction}
It has been widely recognized that three-dimensional (3D) Ising spin glass
(SG) with short range interactions undergoes a second order SG transition
at least at zero magnetic field in thermal equilibrium at a finite spin
freezing temperature $T_{c}$.\cite{Young1997} In spite of enormous amount
of works, however, no consensus has been reached on the nature of the low
temperature SG phase.  Two different pictures are proposed: mean-field (MF)
picture\cite{Mezard1987} and droplet picture.\cite{Fisher1988,Bray1987} In
the MF picture, the low temperature SG phase exhibits a replica symmetry
breaking.  The SG ordered phase consists of infinite number of pure states
organized in an ultrametric structure.  In the droplet picture, in
contrast, the SG phase is described by only two pure thermodynamic states
related to each other by a global spin flip.

The stability of the Ising SG phase in an external magnetic field ($H$) has
been one of the central issues.  In the MF picture, the phase transition
can survive in the presence of low $H$, forming a critical line, so-called
the de Almeida-Thouless (AT) line in the ($H,T$) phase diagram:
\begin{equation} 
H(T)=A(1-T/T_{c})^{p},
\label{eq01} 
\end{equation} 
where $A$ is a field amplitude and the exponent $p=3/2$.\cite{AT1978} This
line separates the paramagnetic (PM) phase from the SG phase.  There is no
change of symmetry at the AT transition.  The correlation length and
relaxation time diverge on crossing this line.  In the droplet picture, no
phase transition occurs in the presence of even infinitesimal $H$ as for a
ferromagnet.  So there is no AT line in the ($H,T$) phase diagram.  Any
apparent transition would be an artifact related to the limited
experimental time scale.  Recent Monte Carlo (MC) simulations on the 3D
Ising Edwards-Anderson (EA)\cite{Edwards1975} model with short range
interactions suggest that the AT line is of dynamic origin and vanishes in
the limit of infinite times.  Takayama,\cite{Takayama2004a} and Takayama
and Hukushima\cite{Takayama2004b} have studied aging phenomena in the 3D
Ising EA model under the $H$-shift aging protocols, using nonequilibrium
Monte Carlo simulations.  They have shown that the characteristic length
scales associated with the $H$-shift aging protocols exhibit unique scaling
behavior.  This scaling behavior implies the instability of the SG phase in
the equilibrium limit even under an infinitesimal $H$.  Young and
Katzgraber\cite{Young2004} have also studied the possibility of the AT line
for the 3D EA Ising SG using MC simulations.  A finite-size scaling of the
correlation length shows no indication of a transition, in contrast to the
zero-field case.  This result also suggests that there is no AT line for
the short range Ising SG. Experimentally Katori and Ito\cite{Katori1994}
have reported the existence of the AT line with $p=1.49$ in the 3D Ising SG
Fe$_{0.5}$Mn$_{0.5}$TiO$_{3}$, supporting the MF picture.  The critical
temperature $T_{g}(H)$ is determined as the one at which zero-field cooled
(ZFC) susceptibility $\chi_{ZFC}$ starts to deviate from the field-cooled
(FC) susceptibility $\chi_{FC}$.  The difference $\Delta\chi = \chi_{FC} -
\chi_{ZFC}$ is a measure for the irreversibility of the system.  Mattson
et al.\cite{Mattson1995} and J\"{o}nsson et al.\cite{Jpnsson2004} have
shown from dynamic scaling analysis for the AC magnetic susceptibility of
the same system that there is no SG transition in the presence of any $H$,
supporting the droplet picture.

Cu$_{0.5}$Co$_{0.5}$Cl$_{2}$-FeCl$_{3}$ graphite bi-intercalation compound
(GBIC) magnetically behaves like a 3D short-ranged Ising
SG.\cite{Suzuki2003,Suzuki2004} This compound undergoes a SG transition at
$T_{c} = 3.92 \pm 0.11$ K in the absence of $H$.  We have determined the
spin freezing temperature $T_{f}(\omega,H)$ at which the differential FC
susceptibility $\partial M_{FC}(T,H)/\partial H$ coincides with the
dispersion $\chi^{\prime}(\omega,T,H=0)$, where $\omega$ ($=2\pi f$) is the
angular frequency.  In the present paper we report experimental result on
the time dependence of $\chi_{ZFC}$ after a ZFC aging protocol with a wait
time $t_{W}$ ($= 1.0 \times 10^{4}$ and $2.0 \times 10^{3}$ sec).  We show
that the relaxation rate $S(t)$ (= d$\chi_{ZFC}(t)$/d$\ln t$) exhibits a
local maximum at a characteristic time $t_{cr}$, reflecting non-equilibrium
aging dynamics.  We find that $t_{cr}$ decreases with increasing $H$ at the
fixed $T$ (2.9 K $\leq T<T_{c}$).  The experimental results of
$T_{f}(\omega,H)$ and $t_{cr}(T,H)$ are discussed in terms of predictions
from both the MF picture and the droplet picture (see Sec.~\ref{back}).  We
show that $T_{f}(\omega,H)$ provides evidence for the instability of the SG
phase in thermal equilibrium in a finite $H$.  We also show that both
$t_{cr}(T,H)$ and $T_{f}(\omega,H)$ exhibit certain scaling behavior,
suggesting a dynamic crossover from SG dynamics to paramagnetic behavior in
the presence of $H$.

\section{\label{back}THEORETICAL BACKGROUND}
\subsection{\label{backA}MF picture: equilibrium SG transition}
In the mean-field picture, each point on the AT line in the $H$-$T$ phase
diagram, $T_{g}(H)$, is a critical point of a continuous phase transition
in the presence of $H$ which exhibits critical divergence of the
correlation length $\xi_{c}(T,H)$ and the correlation time $\tau_{c}(T,H)$
on approaching $T_{g}(H)$ from the high-temperature side.
\begin{equation} 
\tau_{c}(T,H)=\tau_{0}^{*}(\xi_{c}(T,H)/L_{0})^{z},
\label{eq02} 
\end{equation} 
or
\begin{equation} 
\tau_{c}(T,H)=\tau_{0}^{*}[T/T_{g}(H)-1]^{-x},
\label{eq03} 
\end{equation} 
where
\begin{equation} 
\xi_{c}(T,H)=L_{0}[T/T_{g}(H)-1]^{-\nu},
\label{eq04} 
\end{equation} 
$\tau_{0}^{*}$ and $L_{0}$ are microscopic units of time and length, and
$z$ and $\nu$ are critical exponents ($x=\nu z$).  From the onset of the
out-of-equilibrium behavior of the AC susceptibility, the freezing
temperature $T_{f}(\omega,H)$ can be extracted, where the probing time $t$
is given by $t=2\pi/\omega$ for the AC susceptibility.  Then the critical
temperature $T_{g}(H)$ is derived from the condition
$\tau_{c}(T_{f},H)=t=2\pi/\omega$,\cite{Jpnsson2004}
\begin{equation} 
T_{g}(H)=T_{f}(\omega,H)/[(2\pi/\omega\tau_{0}^{*})^{-1/x}+1].
\label{eq05} 
\end{equation} 
If the AT line exists in the system, it is required that $T_{g}(H)$ can be
uniquely determined using the appropriate values of $\tau_{0}^{*}$ and $x$. 
Note that $\tau_{0}^{*}$ and $x$ are assumed to be the microscopic time and
the dynamic critical exponent at $H = 0$, respectively.  The line
$T_{g}(H)$, which is independent of $\omega$, is assumed to form a AT line
of the second order in the $H$-$T$ plane, where $T_{g}(H=0)=T_{c}$.

\subsection{\label{backB}Droplet model: scaling form for aging dynamics}
The situation is rather different in the the droplet model.  After the SG
system is cooled to $T$ ($<T_{c}$) through the ZFC aging protocol, the size
of domain defined by $R_{T}(t_{a})$ grows with the age $t_{a}$ from
$t_{a}=0$ and reaches $R_{T}(t_{W})$ just before the field $H$ is turned on
at the observation time $t=t_{a}-t_{W}=0$ (or
$t_{a}=t_{W}$).\cite{Takayama2004b} Here $R_{T}(t_{a})$ is expressed by
\begin{equation} 
R_{T}(t_{a})=L_{T}(t_{a}/\tau_{0}^{*})^{bT/T_{c}},
\label{eq06} 
\end{equation} 
where $L_{T}$ is a characteristic length, and $b$ is
an exponent.  The aging behavior in $\chi_{ZFC}$ is observed as a function
of $t$.  After $t=0$, a probing length $R_{tr}(t)$ corresponding to the
maximum size of excitation grows with $t$, in a similar way as
$R_{T}(t_{a})$.  Here $R_{tr}(t)$ is the mean size of the subdomain in the
transient regime and is expressed as\cite{Takayama2004b}
\begin{equation} 
R_{tr}(t)=L_{T}(t/\tau_{0}^{*})^{bT/T_{c}}.
\label{eq07} 
\end{equation} 
The quasi equilibrium relaxation occurs first through local spin
arrangements in length $R_{tr}(t) \ll R_{T}(t_{W})$, followed by
non-equilibrium relaxation due to domain growth, when $R_{tr}(t) \approx
R_{T}(t_{W})$, so that a crossover between the short-time quasi-equilibrium
relaxation and the non-equilibrium relaxation at longer observation times
is expected to occur near $t \approx t_{W}$.  In the time range
$0<t<t_{cr}$, which we call the transient regime of $H$-shift processes, SG
domains in local equilibrium with respect to $H>0$ grow within the SG
domains which were in local equilibrium with respect to $H = 0$ at $t_{a} =
t_{W}$.

Here we introduce two characteristic lengths $R_{W}$ and $R_{cr}$ defined
by
\begin{subequations} 
    \label{eq08}
\begin{eqnarray}
R_{W}=R_{T}(t_{a}=t_{W})=L_{T}(t_{W}/\tau_{0}^{*})^{bT/T_{c}}, \label{eq08a}
\\
R_{cr}=R_{tr}(t=t_{cr})=L_{T}(t_{cr}/\tau_{0}^{*})^{bT/T_{c}} , \label{eq08b}
\end{eqnarray}
and the magnetic field crossover length $R_{H}$ defined by\cite{Jpnsson2004}
\begin{equation}
R_{H} \approx l_{H}[(1-T/T_{c})^{a_{eff}}H^{-1}]^{\delta}, 
\label{eq08c}
\end{equation}
\end{subequations}
where $l_{H}$ is a constant associated with a characteristic length,
$\delta =(d/2-\theta)^{-1}$ and $a_{eff}=\theta\nu-\beta/2$.  The
definition of $d$, $\theta$, $\beta$, and $\nu$ is given in the previous
paper.\cite{Suzuki2003} According to Takayama and
Hukushima,\cite{Takayama2004b} it is predicted that $Y$ ($=R_{cr}/R_{H}$)
is described by a scaling function of $X$ ($= R_{W}/R_{H}$):
\begin{equation}
Y=R_{cr}/R_{H}=f(R_{W}/R_{H})=f(X).
\label{eq09} 
\end{equation} 
The $H$-shift aging process is nothing but a dynamic crossover from the SG
phase in $H = 0$ to the paramagnetic state in $H>0$ in the equilibrium
limit.  In the limit of large $R_{W}/R_{H}$, $R_{cr}/R_{H}$ becomes
constant.

For the AC magnetic susceptibility measurement with the angular frequency
$\omega$, the size $R_{tr}(t=2\pi/\omega)$ is the mean size of droplets
that can respond to the AC field of frequency $\omega$.  Then the onset of
the SG out-of-equilibrium dynamics, $T_{f}(t,H)$, can be related to
$R_{H}$ by the condition,\cite{Jpnsson2004}
\begin{subequations} 
    \label{eq10}
\begin{equation}
R_{H}= R_{tr}(t=2\pi/\omega), \label{eq10a}
\end{equation}
or
\begin{equation}
(1-T_{f}(\omega,H)/T_{c})^{a_{eff}}/H =(L_{T}/l_{H})^{1/\delta}
[(2\pi/\omega \tau_{0}^{*})^{T_{f}(t,H)/T_{c}}]^{b/\delta}. \label{eq10b}
\end{equation}
\end{subequations}

\section{\label{exp}EXPERIMENTAL PROCEDURE}
The detail of sample characterization and sample preparation of
Cu$_{0.5}$Co$_{0.5}$Cl$_{2}$-FeCl$_{3}$ GBIC was provided in our previous
papers.\cite{Suzuki2003,Suzuki2004} The DC magnetization was measured using
a SQUID magnetometer (Quantum Design, MPMS XL-5) with an ultra low field
capability option.  The remnant magnetic field was reduced to zero field
(exactly less than 3 mOe) at 298 K. The time ($t$) dependence of the
zero-field cooled (ZFC) magnetization ($M_{ZFC}$) was measured at fixed
points in the ($H,T$) plane.  The following ZFC aging protocol was made
before the measurement.  First the sample was annealed at 50 K for $1.2
\times 10^{3}$ sec in the absence of $H$.  Then the system was quenched
from 50 K to $T$ ($<T_{c}$).  It was kept at $T$ for a wait time $t_{W}$
($= 2.0 \times 10^{3}$ sec and $1.0 \times 10^{4}$ sec).  After the wait
time, an external magnetic field $H$ is applied along any direction
perpendicular to the $c$-axis at $t = 0$ (or $t_{a} = t_{W}$), where
$t_{a}$ is an age and $t$ is an observation time ($t_{a} = t + t_{W}$). 
The measurements of $\chi_{ZFC}$ vs $t$ were carried out at the fixed $T$
($2.9 \leq T < T_{c}$) when $H$ is varied as a parameter: $1 \leq H \leq
400$ Oe.

Under the above conditions, $\chi_{ZFC}(t)$ ($=M_{ZFC}(t)/H$) was measured
as a function of $t$.  The relaxation rate defined by $S(t)$ (=
d$\chi_{ZFC}(t)$/d$\ln t$) exhibits a peak at a characteristic time
$t_{cr}$.\cite{Lundgren1990} Theoretically\cite{Ogielski1985,Koper1988} and
experimentally\cite{Hoogerbeets1985,Alba1986,Suzuki2005} it has been
accepted that the time variation of $\chi_{ZFC}(t)$ may be described by a
stretched exponential relaxation form:
\begin{equation}
\chi_{ZFC}(t) = \chi_{0}-B_{0}t^{-m}\exp[-(t/\tau)^{1-n}],
\label{eq11}
\end{equation}
where the exponent $m$ may be positive and is very close to zero, $n$ is
between 0 and 1, $\tau$ is a characteristic relaxation time, $\chi_{0}$,
$m$, and $B_{0}$ are constants, and $n$ is a stretched exponential
exponent.  The least-squares fit of the data of $\chi_{ZFC}$ vs $t$ to
Eq.(11) with $m = 0$ yields the parameters $n$, $\tau$, and $B_{0}$ as a
function of $H$ and $T$ in the $H$-$T$ plane.

\section{\label{result}RESULT}
\subsection{\label{resultA}Line $T_{g0}(H)$ from the measurement of
d$\Delta\chi$/d$T$ vs $T$}

\begin{figure}
\includegraphics[width=8.0cm]{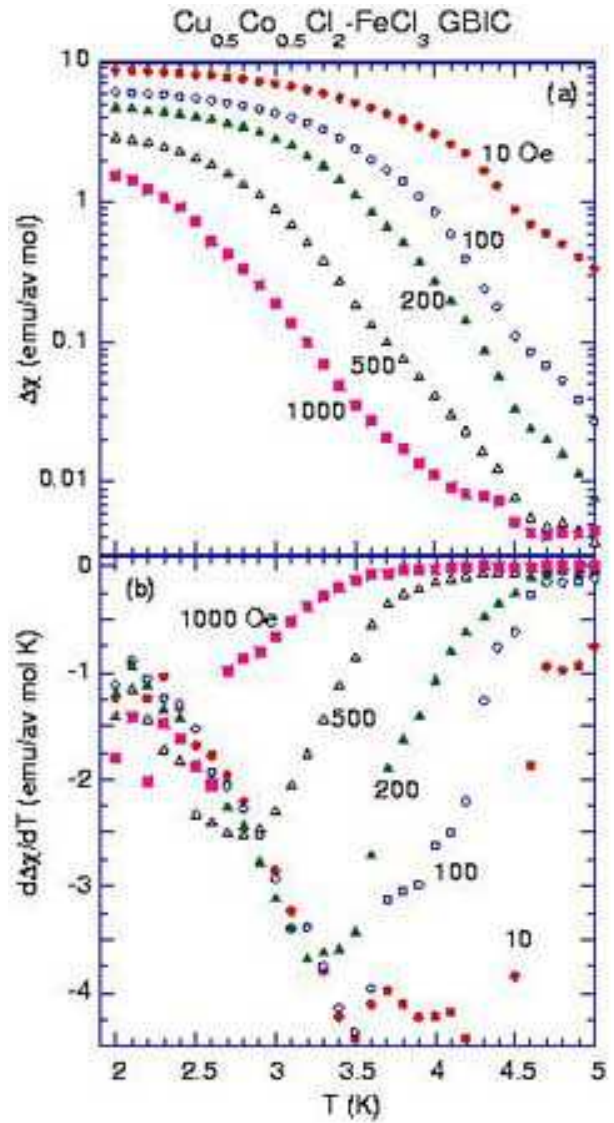}
\caption{\label{fig01}(Color online) $T$ dependence of (a) $\Delta\chi =
\chi_{FC} - \chi_{ZFC}$ and (b) d$\Delta\chi$/d$T$ at various $H$ for
Cu$_{0.5}$Co$_{0.5}$Cl$_{2}$-FeCl$_{3}$ GBIC.}
\end{figure}

\begin{figure}
\includegraphics[width=8.0cm]{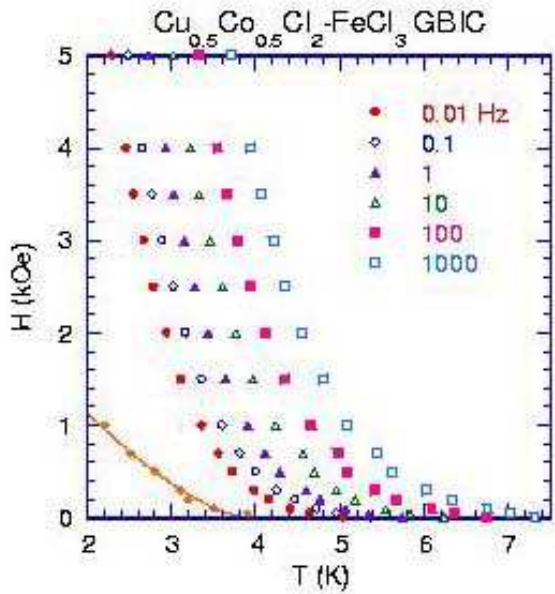}
\caption{\label{fig02}(Color online) $H$-$T$ phase diagram.  (i) The local
minimum temperature of d$\Delta\chi$/d$T$ vs $T$ at $H$ 
($\blacklozenge$) [denoted as the line $T_{g0}(H)$ (d$\Delta\chi$/d$T$ vs $T$)]. 
A solid line is a least-squares fitting curve which is expressed by
Eq.(\ref{eq01}) with $T_{c}$, $p$ and $A$ given in the text.  (ii)
$T_{f}(\omega,H)$ vs $H$ between $f = 0.01$ Hz and 1 kHz.
$T_{f}(\omega,H)$ is defined as a temperature at which $\chi_{FC}(T,H)$
crosses $\chi^{\prime}(\omega,T,H=0)$ at each $\omega$.}
\end{figure}

In our previous paper\cite{Suzuki2003} we have discussed the $H$ dependence
of the spin freezing temperature in detail.  The result is summarized as
follows.  Figure \ref{fig01}(a) shows the $T$ dependence of the difference
between $\chi_{FC}$ and $\chi_{ZFC}$ ($\Delta\chi = \chi_{FC}-\chi_{ZFC}$)
at various $H$ in the vicinity of $T_{c}$.  The transition temperature is
defined as a temperature at which $\Delta\chi = 0$.  Since it is difficult
to determine exactly the transition temperature from Fig.~\ref{fig01}(a),
we redefine the line $T_{g0}(H)$ as one at which d$\Delta\chi$/d$T$ vs $T$
shows a local minimum for each $H$ (see Fig.~\ref{fig01}(b)).  In
Fig.~\ref{fig02} we show the $H$ dependence of $T_{g0}(H)$ thus obtained. 
The least squares fit of the data of $H$ vs $T$ (or $H$ vs $T_{g0}(H)$ in
Fig.~\ref{fig02} to Eq.(\ref{eq01}) yields the parameters $p$ and $A$,
where $T_{c}$ is fixed as $T_{c} = 3.92 \pm 0.11$ K. The exponent $p$ is
equal to $1.52 \pm 0.10$, which is close to the AT exponent ($p=3/2$).  The
value of $A$ is $A_{exp} = 3.38 \pm 0.67$ kOe.  For convenience, hereafter
this critical line is denoted as the line $T_{g0}(H)$ (d$\Delta\chi$/d$T$
vs $T$).  In the classical $q$ vector SG model, the value of $A$ is
predicted as\cite{Hoogerbeets1985,Chu1995}
\begin{equation}
A_{theory} = [8/(q+1)(q+2)]^{1/2}(k_{B}T_{c}/g\mu_{B}S),
\label{eq12}
\end{equation}
where $S$ is the spin, $g$ is the Land\'{e} $g$-factor, and $q$ is the
number of spin degrees of freedom ($q=1$ for Ising SG and $q=3$ for
Heisenberg SG).  Here we assume $g =2$.  The value of $S$ is estimated as
$S =1.56$ for our system with the stoichiometry
C$_{5.26}$(Cu$_{0.5}$Co$_{0.5}$Cl$_{2}$)$_{0.47}$(FeCl$_{3}$)$_{0.53}$,
\cite{Suzuki2003} where $S = 1/2$ for Co$^{2+}$ and Cu$^{2+}$ and $S = 5/2$
for Fe$^{3+}$.  Then the value of $A_{theory}$ is estimated as 21.6 kOe for
$q = 1$ (Ising symmetry), leading to $A_{exp}/A_{theory} = 0.16$.  These
results are similar to those reported by Katori and Ito\cite{Katori1994}
for Fe$_{0.5}$Mn$_{0.5}$TiO$_{3}$: $p = 1.49$ and $A_{exp}/A_{theory} =
0.52$ ($A_{exp} = 42.6$ kOe, $A_{theory} = 82.1$ kOe).  Although such
result apparently supports the AT theory, it does not provide any evidence
for an equilibrium SG transition in the presence of $H$.

\subsection{\label{resultB}Line $T_{f}(\omega,H)$ where $\partial 
M_{FC}(T,H)/\partial H = \chi^{\prime}(\omega,T,H=0)$}
In our previous paper\cite{Suzuki2003} we have discussed the spin freezing
temperature $T_{f}(\omega,H)$ of Cu$_{0.5}$Co$_{0.5}$Cl$_{2}$-FeCl$_{3}$,
at which $\partial M_{FC}(T,H)/\partial H$ crosses the dispersion
$\chi^{\prime}(\omega,T,H=0)$ at each $f$.  Such definition of
$T_{f}(\omega,H)$ has been also used by Mattsson et al.\cite{Mattson1995}
and J\"{o}nsson et al.\cite{Jpnsson2004} for $T_{f}(\omega,H)$ in
Fe$_{0.5}$Mn$_{0.5}$TiO$_{3}$.  In Fig.~\ref{fig02}, we show the plot of
$T_{f}(\omega,H)$ as a function of $H$ where $f$ ($0.01 \leq f \leq 1000$
Hz) is varied as a parameter.  The line $T_{f}(\omega,H)$ is strongly
dependent on $f$.  If $T_{f}(\omega,H)$ In the limit of $f = 0$ coincides
with the line $T_{g0}(H)$ (d$\Delta\chi$/d$T$ vs $T$) as shown in
Fig.~\ref{fig02}, then this line would be an equilibrium AT line.

It has been reported that the SG phase transition occurs only at zero field
in Fe$_{0.5}$Mn$_{0.5}$TiO$_{3}$.\cite{Mattson1995,Jpnsson2004} The
disapperance of the AT line at finite fields is consistent with the
prediction derived from the droplet picture.  By using the same method (see
Sec.~\ref{back}), here we examine whether the system exhibits the critical
slowing down in the presence of $H$. If the SG transition occurs at
$T_{g}(H)$ in thermal equilibrium as is predicted from the MF picture, it
is required that the critical temperature $T_{g}(H)$ evaluated from
Eq.(\ref{eq05}) should fall on a unique curve $T$ = $T_{g}(H)$, which is
independent of $\omega$.

\begin{figure}
\includegraphics[width=8.0cm]{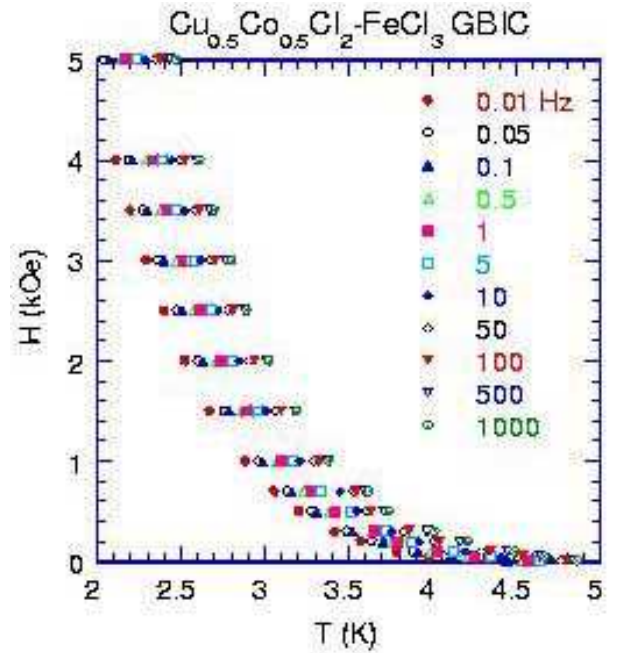}
\caption{\label{fig03}(Color online) Plot of $T_{g}(H)$ ($=
T_{f}(\omega,H)/[1 + (2\pi$/$\omega\tau_{0}^{*})^{-1/x}$]) vs $H$ with $x$ =
10.3 and $\tau_{0}^{*} = 5.27 \times 10^{-6}$ sec, where $\omega$ ($= 2\pi
f$) is the angular frequency.}
\end{figure}

We calculate the line $T_{g}(H)$ using Eq.(\ref{eq05}) with the data of
$T_{f}(\omega,H)$ for each $\omega$.  Here we use $\tau_{0}^{*}=5.27 \times
10^{-6}$ sec, $T_{c} = 3.92$ K, and $x = z\nu = 10.3$, which are obtained
from the analysis of the absorption $\chi^{\prime\prime}(\omega,T,H=0)$ in
the previous paper.\cite{Suzuki2003} Figure \ref{fig03} shows the line
$T_{g}(H)$ thus obtained.  We find that the line $T_{g}(H)$ is still
strongly dependent on $\omega$, although the line $T_{g}(H)$ tends to
approach the line $T_{g0}(H)$ (d$\Delta\chi$/d$T$).  We note that the line
$T_{g}(H)$ is still dependent of $\omega$ even if $x$ is equal to 20, which
is unphysically large.  This result suggests that no equilibrium SG phase
transition occurs in the presence of $H$, which is inconsistent with the
prediction from the MF picture.

\begin{figure}
\includegraphics[width=8.0cm]{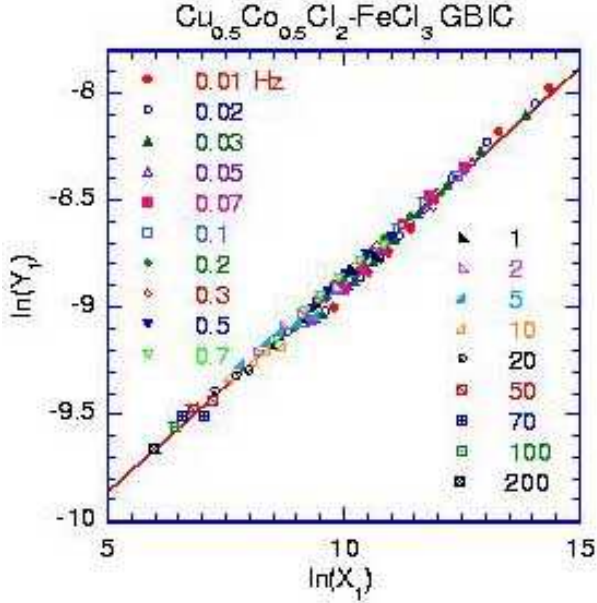}
\caption{\label{fig04}(Color online) Scaling plot of $\ln(Y_{1})$ vs
$\ln(X_{1})$ for the data with $T_{f}(\omega,H)/T_{c} \leq 0.9$, where
$Y_{1}=(1-T_{f}(\omega ,H)/T_{c})^{a_{eff}}/H$ and
$X_{1}=(2\pi/\omega\tau_{0}^{*})^{T_{f}(t,H)/T_{c}}$.  The best result for
the scaling plot is obtained when $a_{eff} = 0.55$, where $T_{c}$ (= 3.92
K) and $\tau_{0}^{*}$ ($= 5.27 \times 10^{-6}$ sec) are fixed.  The
least-sqaures fitting curve ($Y_{1} \approx \zeta^{1/\delta}
X_{1}^{b/\delta}$) is denoted by a straight line with a slope of $b/\delta
= 0.194$ and $\zeta = 1.5 \times 10^{-4}$ ($\delta = 0.81$).}
\end{figure}

Now we examine the validity of the droplet picture.  As is described in
Sec.~\ref{back}, the onset of the SG out-of-equilibrium dynamics,
$T_{f}(t,H)$, is related to $R_{H}$ by Eq.(\ref{eq10b}) with
$X_{1}=(t/\tau_{0}^{*})^{T_{f}(t,H)/T_{c}}$ and
$t=2\pi/\omega$.\cite{Jpnsson2004} When $Y_{1}$ is defined as $Y_{1}
=(1-T_{f}(t,H)/T_{c})^{a_{eff}}/H$, Eq.(\ref{eq10b}) is rewritten as
\begin{equation}
Y_{1} \approx \zeta^{1/\delta}X_{1}^{b/\delta },
\label{eq13}
\end{equation}
with $\zeta$ = $L_{T}$/$l_{H}$.  Figure \ref{fig04} shows the plot of
$\ln(Y_{1})$ as a function of $\ln(X_{1})$ with the best choice of $a_{eff}
= 0.55 \pm 0.05$, where $\tau_{0}^{*} = 5.27 \times 10^{-6}$ sec.  When ony
the data with $T_{f}(t,H)/T_{c} < 0.9$ are used, almost all the data fall
well on an unique straight line with the slope $b/\delta = 0.198 \pm 0.002$
and $\ln(\zeta^{1/\delta}) = -10.85 \pm 0.02$.  Using the value of $b$ =
0.16 for $\chi^{\prime\prime}(\omega,T= 3.75$ K$,t)$ with $f = 0.05$
Hz,\cite{Suzuki2003} $\delta$ can be estimated as $\delta = 0.81$.  This
value of $\delta$ is slightly deviated from the calculated value of
$\delta$ using the relation $\delta = (d/2 - \theta)^{-1} = 0.73$ with $d =
3$ and $\theta = 0.13$.  Our value of $a_{eff}$ is much larger than the
value predicted from the relation ($a_{eff} = \theta\nu - \beta/2 \approx
0$), where $\beta = 0.36 \pm 0.03$, $\nu = 1.4 \pm 0.2$, and $\theta = 0.13
\pm 0.02$ which are obtained in the previous paper.\cite{Suzuki2003} Using
$\delta = 0.81$, the value of $\zeta$ is calculated as $(1.5 \pm 0.1)
\times 10^{-4}$.  This value will be used in Sec.~\ref{dis}.  It is
interesting to compare our result with that reported by J\"{o}nsson et al. 
for Fe$_{0.5}$Mn$_{0.5}$TiO$_{3}$.\cite{Jpnsson2004} They have reported the
values of $a_{eff} = 0.45$ and $b/\delta = 0.17$ for the high-temperature
(HT) data with $T_{f}(\omega,H)/T_{c} \leq 0.95$ and $H \leq 5$ kOe, and
$a_{eff} = 0.25$, $b/\delta = 0.143$ for the low-temperature (LT) data,
where $\theta = 0.2$ and $\delta = 0.77$.  Our result ($a_{eff} = 0.55$ and
$b/\delta = 0.198$) is close that obtained from the HT data.

\subsection{\label{resultC}Stretched exponetial relaxation form
$\chi_{ZFC}(t)$ at $H=1$ Oe}

\begin{figure}
\includegraphics[width=8.0cm]{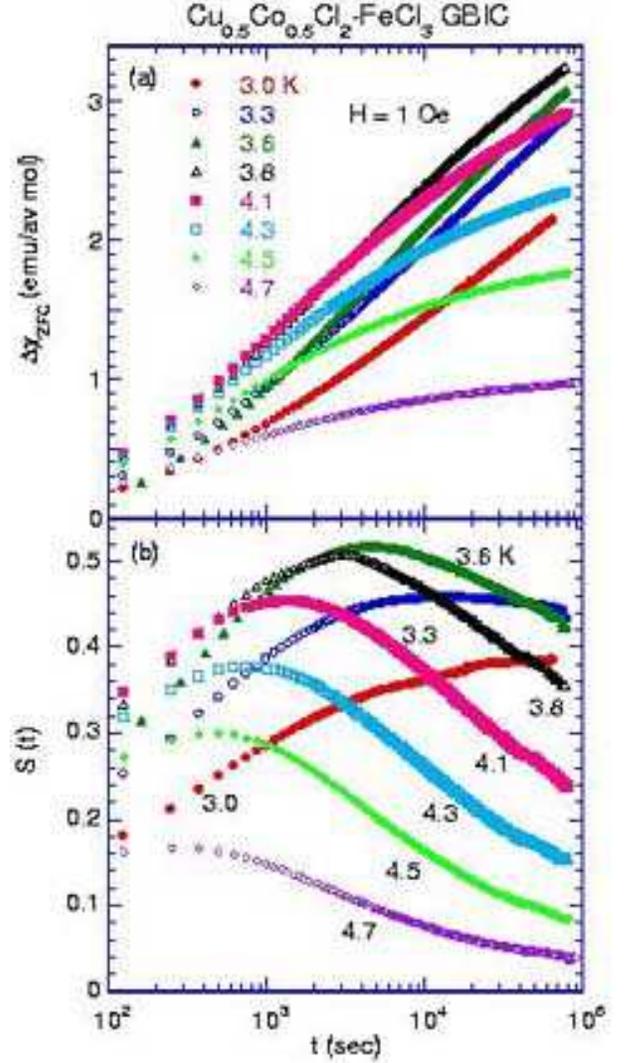}
\caption{\label{fig05}(Color online) $t$ dependence of (a)
$\Delta\chi_{ZFC}(t)$ [$= \chi_{ZFC}(t) - \chi_{ZFC}(t=0)$] and (b) the
relaxation rate $S(t)$ (= d$\chi_{ZFC}(t)$/d$\ln t$) at various $T$ ($3.0
\leq T \leq 4.7$ K).  $t_{W} = 2.0 \times 10^{3}$ sec.  $H =1$ Oe.}
\end{figure}

\begin{figure}
\includegraphics[width=8.0cm]{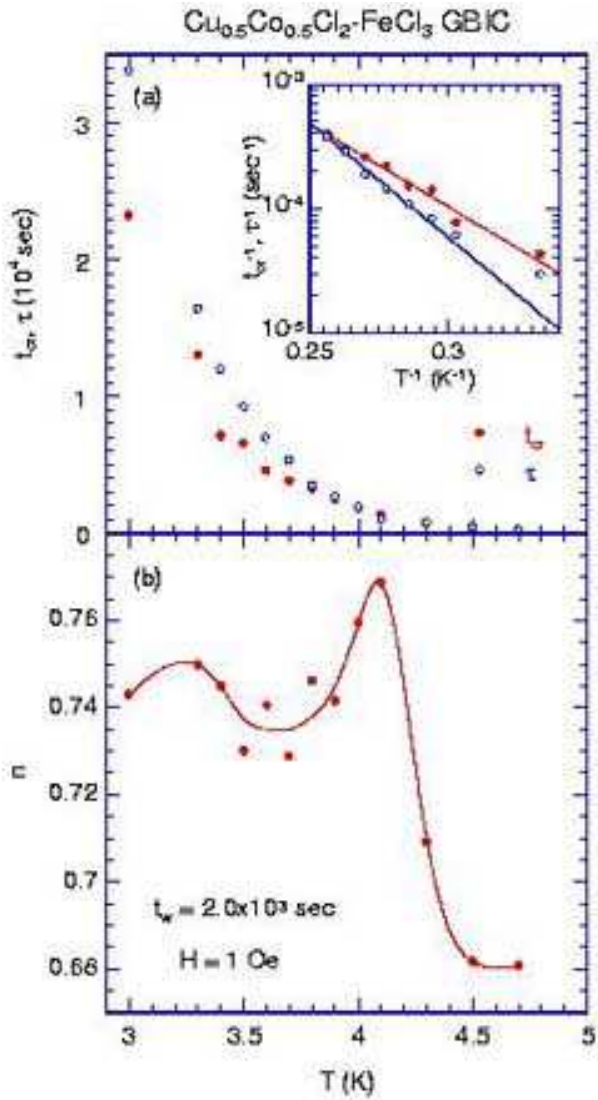}
\caption{\label{fig06}(Color online) (a) $T$ dependence of $t_{cr}$ and
$\tau$.  $t_{cr}$ is a characteristic time at which $S(t)$ exhibits a peak
and $\tau$ is the relaxation time for the stretched exponential relaxation. 
$H = 1$ Oe.  $t_{W} = 2.0 \times 10^{3}$ sec.  The inset shows the plots of
$1/t_{cr}$ vs $1/T$ and $1/\tau$ vs $1/T$ for $T<T_{c}$.  The solid lines
are least-squares fits to Eq.(\ref{eq14}) for both $1/t_{cr}$ vs $1/T$ and
$1/\tau$ vs $1/T$.  The fitting parameters are given in the text.  (b) $T$
dependence of the stretched exponential exponent $n$.  The solid line is a
guide to the eyes.}
\end{figure}

We have measured the $t$ dependence of $\chi_{ZFC}(t)$ after the ZFC aging
protocol described in Sec.~\ref{exp}.  Figures \ref{fig05}(a) and (b) show
the $t$ dependence of $\chi_{ZFC}(t)$ and the relaxation rate $S(t)$ 
at various $T$, respectively, where $t_{W} = 2.0
\times 10^{3}$ sec and $H = 1$ Oe.  The relaxation rate $S(t)$ exhibits a
peak at $t = t_{cr}$, which shifts to the short-$t$ side with increasing
$T$.  The ZFC susceptibility $\chi_{ZFC}(t)$ is well described by
Eq.(\ref{eq11}) with $m = 0$ for $1.0 \times 10^{2} < t < 6.0 \times
10^{4}$ sec.  The least-squares fit of the data of $\chi_{ZFC}$ vs $t$
yields the parameters $n$ and $\tau$ for each $T$.  In Fig.~\ref{fig06}(a),
we show the $T$ dependence of $t_{cr}$ and $\tau$.  Both $t_{cr}$ and
$\tau$ increase with decreasing $T$: $\tau = 3.4 \times 10^{4}$ sec at $T =
3$ K. Note that $\tau$ is larger than $t_{cr}$ below $T_{c}$.  In the inset
of Fig.~\ref{fig06}(a) we show the plot of $1/t_{cr}$ and $1/\tau$ as a
function of $1/T$.  The least-squares fit of the data of $1/\tau$ vs $1/T$
for $T<T_{c}$ to the form\cite{Hoogerbeets1985}
\begin{equation}
1/\tau = c_{1}\exp(-c_{2}T_{c}/T),
\label{eq14}
\end{equation}
yields the parameters $c_{1} = 19.7 \pm 11.5$ sec$^{-1}$ and $c_{2} = 10.82
\pm 0.56$.  For the data of $1/t_{cr}$ vs $1/T$, similarly we obtain $c_{1}
= 1.01 \pm 0.48$ sec$^{-1}$ and $c_{2} = 7.81 \pm 0.46$.  Note that our
value of $c_{2}$ is much larger than those derived by Hoogerbeets et
al.\cite{Hoogerbeets1985} ($c_{2} = 2.5$) from their analysis of
thermoremnant magnetization (TRM) relaxation measurements on four canonical
SG systems: Ag: Mn (2.6 at.~\%), Ag: Mn (4.1 at.~\%), Ag: [Mn (2.6 at.~\%)
+ Sb (0.46 at.~\%)], and Cu: Mn (4.0 at.~\%).  Note that in their work the
stretched exponential is taken as representative of the short time
($t<t_{W}$) relaxation.  In Fig.~\ref{fig06}(b) we show the $T$ dependence
of $n$.  The exponent $n$ exhibits a local maximum around $T = 3.3$ K
($T/T_{c} = 0.84$), a local minimum around 3.5 - 3.7 K ($T/T_{c} = 0.89 -
0.94$), and a sharp peak just above $T_{c}$.  It drastically decreases with
further increasing $T$ above $T_{c}$.  The overall variation of $n$ below
$T_{c}$ is similar to that reported by Bontemps\cite{Bontemps1986} and
Bontemps and Orbach\cite{Bontemps1988} for Eu$_{0.4}$Sr$_{0.6}$S ($T_{c} =
1.50$ K): $n$ has a local maximum at $T/T_{c} = 0.87$, a local minimum at
$T/T_{c} = 0.91$, and a local maximum at $T = T_{c}$.  In summary, the
transition between the SG and PM phases at $H=1$ Oe is characterized by the
peak of $n$ and the drastic increase of $t_{cr}$ (or $\tau$) with
decreasing $T$ around $T_{c.}$

\subsection{\label{resultD}$H$ and $T$ dependence of $t_{cr}$ and $\tau$}

\begin{figure}
\includegraphics[width=8.0cm]{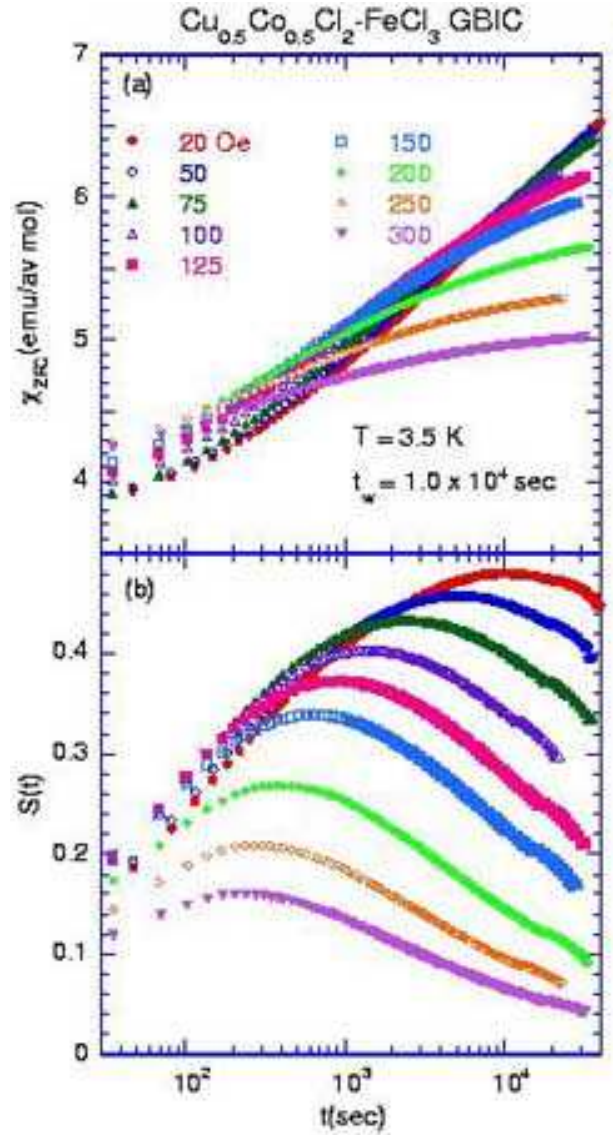}
\caption{\label{fig07}(Color online) $t$ dependence of (a) $\chi_{ZFC}(t)$
and (b) $S(t)$ at various $H$ ($20 \leq H \leq 300$ Oe).  $t_{W} = 1.0
\times 10^{4}$ sec.  $T = 3.50$ K.}
\end{figure}

\begin{figure}
\includegraphics[width=8.0cm]{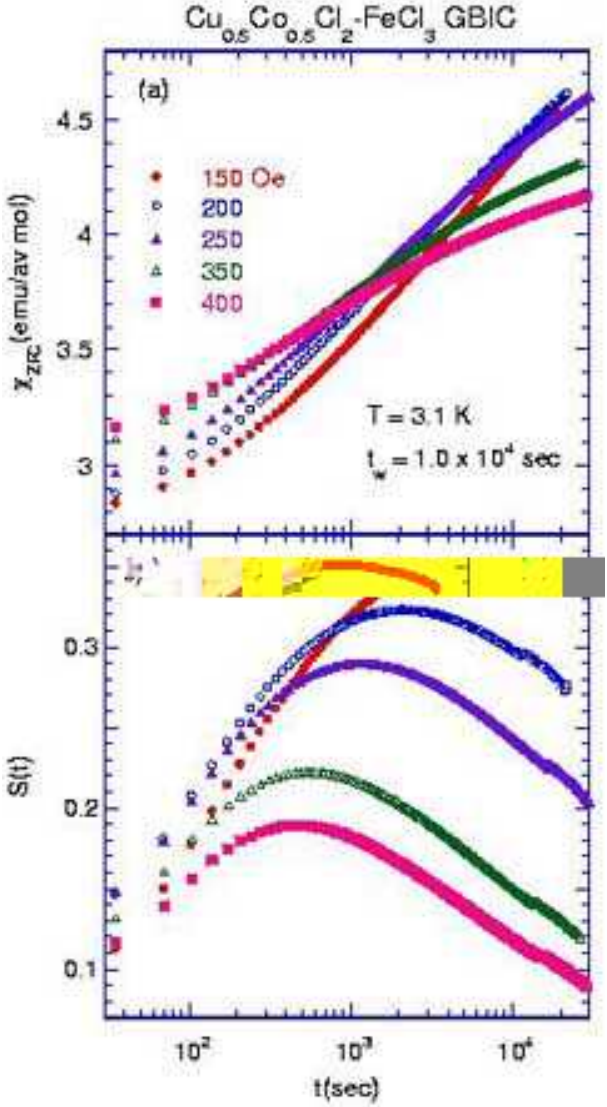}
\caption{\label{fig08}(Color online) $t$ dependence of (a) $\chi_{ZFC}(t)$
and (b) $S(t)$ at various $H$ ($150 \leq H \leq 400$ Oe).  $t_{W} = 1.0
\times 10^{4}$ sec.  $T = 3.10$ K.}
\end{figure}

We have also measured the $t$ dependence of $\chi_{ZFC}(t)$ after the same
ZFC aging protocol described in Sec.~\ref{exp}, where $t_{W} = 1.0\times
10^{4}$ sec, $T$ = 2.9, 3.0, 3.1, 3.3, 3.5, 3.65, and 3.75 K and $H$ = 20,
50, 100, 150, 200, 300, and 400 Oe.  Figures \ref{fig07}(a) and (b) show
the $t$ dependence of $\chi_{ZFC}(t)$ and $S(t)$ at $T = 3.5$ K,
respectively.  Figures \ref{fig08}(a) and (b) show the $t$ dependence of
$\chi_{ZFC}(t)$ and $S(t)$ at $T = 3.1$ K, respectively.  Here $H$ is
chosen such that the point at ($T,H$) crosses the line $T_{g0}(H)$
(d$\Delta\chi$/d$T$ vs $T$) in the $H$-$T$ plane.  The relaxation $S(t)$
has a peak at $t=t_{cr}$, which shifts to the low-$t$ side with increasing
$H$.  The peak height of $S(t)$ at $t=t_{cr}$ drastically decreases with
increasing $H$.  The least-squares fit of the data of $\chi_{ZFC}(t)$ vs
$t$ for $1.0 \times 10^{2}$ sec $<t<t_{W}$ to Eq.(\ref{eq11}) with $m = 0$
yields the exponent $n$ and $\tau$ at each $H$ and $T$, where $\tau$ is the
relaxation time of the stretched exponential relaxation.  Note that the
data ($\chi_{ZFC}$ vs $t$) are not fitted well with Eq.(\ref{eq11}) with $m
\neq 0$: the value of $n$ is dependent on the small change in $m$.

\begin{figure}
\includegraphics[width=8.0cm]{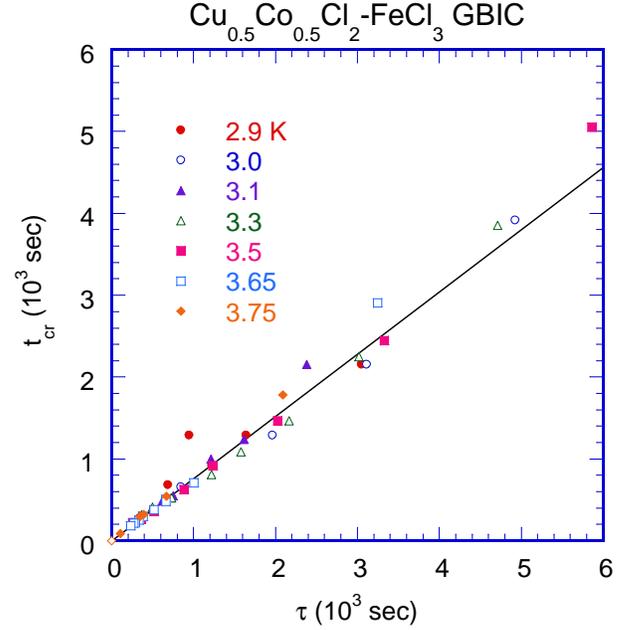}
\caption{\label{fig09}(Color online) Plot of $t_{cr}$ vs $\tau$ for various
$H$ between 50 and 400 Oe at each $T$.  A solid line is a least-squares
fitting curve denoted by $x_{cr} = t_{cr}/\tau = 0.80$. $t_{W}=1.0 
\times 10^{4}$ sec.}
\end{figure}

\begin{figure}
\includegraphics[width=8.0cm]{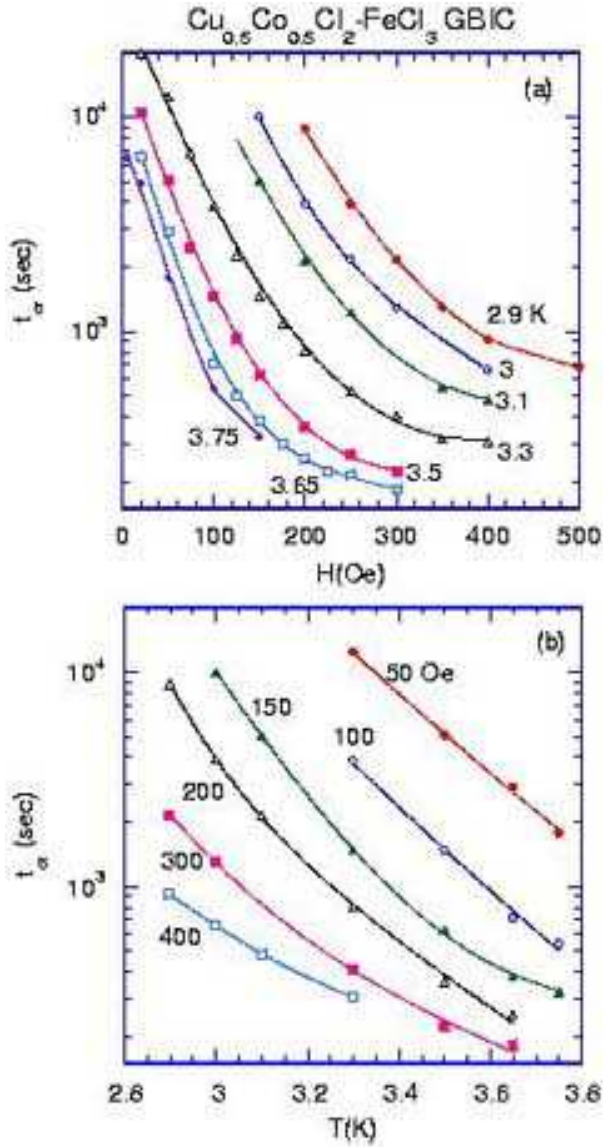}
\caption{\label{fig10}(Color online) (a) $t_{cr}$ vs $H$ at various $T$ and
(b) $t_{cr}$
vs $T$ at various $H$.  The solid lines are guides to the eyes.}
\end{figure}

The values of $\tau$ and $t_{cr}$ have been determined from the $t$
dependence of $\chi_{ZFC}(t)$ and $S(t)$ at fixed $H$ and $T$.  In
Fig.~\ref{fig09} we make a plot of $t_{cr}$ vs $\tau$ at each $T$ ($2.9
\leq T \leq 3.75$ K) for measured values of $H$.  We find that $t_{cr}$ is
proportional to $\tau$ in the range $0 \leq \tau \leq 6.0 \times 10^{3}$
sec: $x_{cr} = t_{cr}/\tau = 0.80 \pm 0.01$.  This ratio is different from
the ratio ($x_{cr} = 1$) predicted from Eq.(\ref{eq11}) with $m = 0$, which
is independent of $n$.  A slight deviation from this linear relation is
observed for $\tau > 6.0 \times 10^{3}$ corresponding to the case of low $H$
($H<50$ Oe), partly because of the larger uncertainty in $\tau$.  Since the
$H$ and $T$ dependence of $t_{cr}$ is essentially the same as that of
$\tau$, hereafter we discuss only the $H$ and $T$ dependence of $t_{cr}$. 
Figure \ref{fig10}(a) shows the $H$ dependence of $t_{cr}$ at fixed $T$
($2.9 \leq T \leq 3.75$ K) below $T_{c}$.  At higher $T$, the decrease of
$t_{cr}$ with $H$ occurs at lower $H$.  Figure \ref{fig10}(b) shows the $T$
dependence of $t_{cr}$ at fixed $H$ below $T_{c}$.

\section{\label{dis}DISCUSSION}

\begin{figure}
\includegraphics[width=8.0cm]{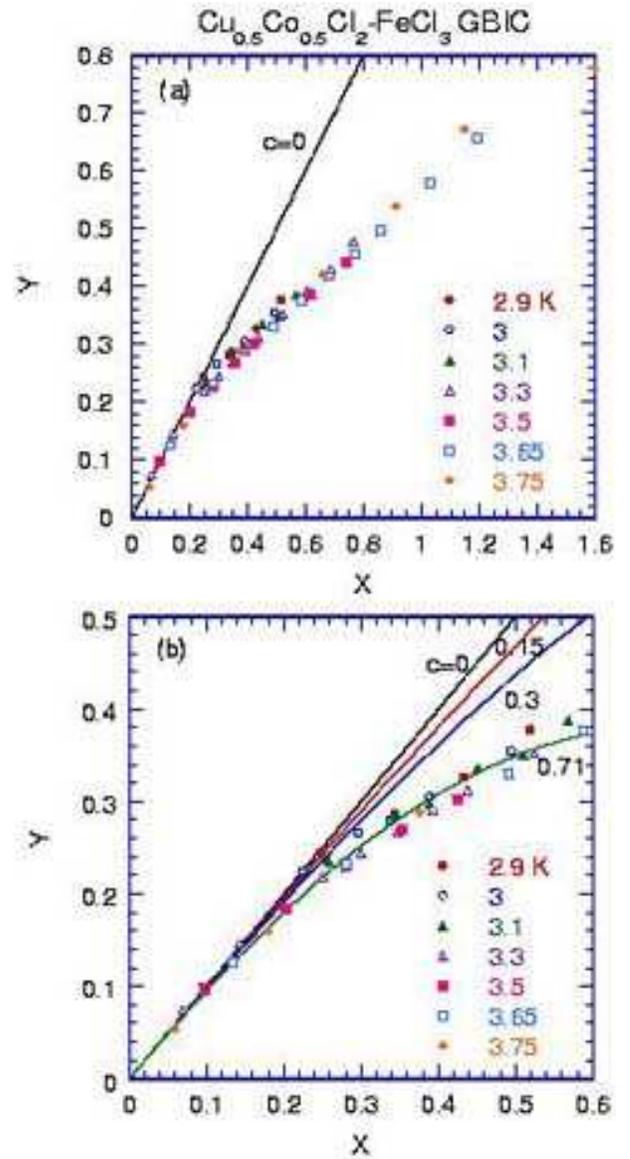}
\caption{\label{fig11}(Color online) (a) Scaling plot of $Y$ vs $X$.  $X$
and $Y$ are defined by Eqs.(\ref{eq15a}) and (\ref{eq15b}), respectively,
where $\delta/b = 5.15$, $b = 0.16$, $a_{eff}=0.55$, $\zeta = 1.5 \times
10^{-4}$, $T_{c} = 3.92$ K, and $\tau_{0}^{*} = 5.27 \times 10^{-6}$ sec. 
$t_{cr}$ is a time of the peak for $S(t)$ vs $t$ with $t_{W} = 1.0 \times
10^{4}$ sec for $2.9 \leq T \leq 3.75$ K and $5 \leq H \leq 500$ Oe.  (b)
The same scaling plot of $Y$ vs $X$ for $0 \leq X \leq 0.6$.  The solid
lines are denoted by Eq.(\ref{eq16}) with $c$ = 0, 0.15, 0.3, and 0.71
where $\delta = 0.81$.  Note that the solid line with $c = 0.71$ also
coincides with that derived from the least-squares fit of the data to
Eq.(\ref{eq16}) with $c = 0.71 \pm 0.06$ and $\delta = 0.81 \pm 0.07$.}
\end{figure}

We examine the validity of the scaling relation given by Eq.(\ref{eq09})
using the data of $t_{cr}(T,H)$ with $t_{W} = 1.0 \times 10^{4}$ sec in
Figs.~\ref{fig10}(a) and (b).  Note that $t_{cr}(T,H)$ is the peak time of
$S(t)$ vs $t$ for $H$ $2.9 \leq T \leq 3.75$ K and $5 \leq H \leq 500$ Oe. 
We define the following two parameters $X$ and $Y$ given by
\begin{subequations}
    \label{eq15}
\begin{equation}
X=\zeta [H^{\delta/b}(t_{W}/\tau_{0}^{*})^{T/T_{c}}
(1-T/T_{c})^{-(\delta /b)a_{eff}}]^{b}, \label{eq15a}
\end{equation}
and 
\begin{equation}
Y=\zeta [H^{\delta /b} (t_{cr}(T,H)/\tau_{0}^{*})^{T/T_{c}}
(1-T/T_{c})^{-(\delta /b)a_{eff}}]^{b}, \label{eq15b}
\end{equation}
\end{subequations}
respectively.  In Figs.~\ref{fig11}(a) and (b) we show a scaling plot of
$Y$ vs $X$ with $\tau_{0}^{*} = 5.27 \times 10^{-6}$ sec and $T_{c} = 3.92$
K. If we use the parameters $\delta/b$ (= 5.05), $a_{eff}$ (= 0.55), and
$\zeta$ ($= 1.5 \times 10^{-4}$), which are determined independently from
Fig.~\ref{fig04}, we find that almost all the points $(X,Y)$ fall
well on a single curve.  Here we use the value of $b$ (=
0.16).\cite{Suzuki2003} In fact, the scaling form of Figs.~\ref{fig11}(a)
and (b) is not so sensitive to the choice of $b$ for $0.14 \leq b \leq
0.18$.  Similar scaling curve has been reported by J\"{o}nsson et
al.\cite{Jonsson2002,Jonsson2003} from the aging behavior of 
$\chi_{ZFC}(t)$ for Ag 11 at.~\% Mn under the $T$-shift protocol.

Our curve is well described by a straight line given by $Y=X$ for $X \leq
0.1$.  It deviates from the straight line with increasing $X$ and tends to
saturate around $X \approx 1.6$.  There is a gradual crossover between
accumulative and chaos around $X = 0.1$.  Our curve is similar to a scaling
curve of $Y$ vs $X$ which is predicted by Takayama and
Hukushima\cite{Takayama2004b} from the Monte Carlo simulation on the 3D
Ising EA model under the $H$-shift protocol.  The scaling curve is
described by
\begin{equation}
Y = X - cX^{1+1/\delta},
\label{eq16}
\end{equation}
for $X \ll 1$, where $c$ is a constant.  In Fig.~\ref{fig11}(b), for
comparison we show the plot of $Y$ vs $X$ given by Eq.(\ref{eq16}) with
$\delta = 0.81$, when $c$ is changed as a parameter.  Our data are fitted
well with Eq.(\ref{eq16}) with $c = 0.15$ for $X \leq 0.3$.  Note that when
$c$ and $\delta$ are adjustable parameters, the least squares fit of the
data ($Y$ vs $X$) ($0 \leq X \leq 0.6$) to Eq.(\ref{eq16}) yields the
parameters $c = 0.71 \pm 0.06$ and $\delta =
0.81 \pm 0.07$.

\begin{figure}
\includegraphics[width=8.0cm]{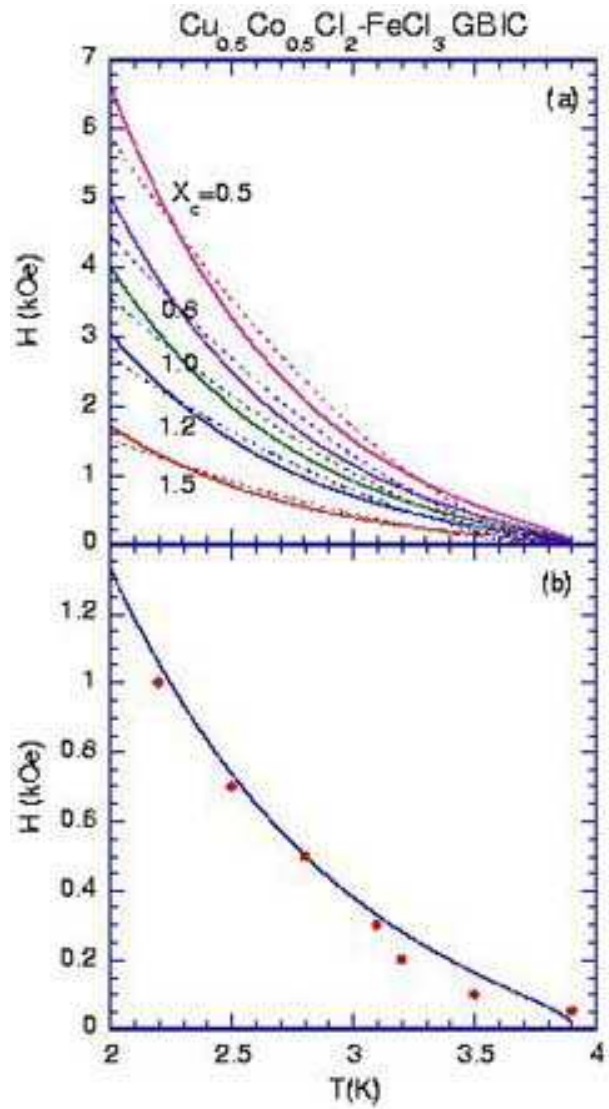}
\caption{\label{fig12}(Color online) (a) $H$-$T$ line (denoted by 
the solid lines).  The value of
$H$ is calculated from Eq.(\ref{eq17}) with $t_{W} = 1.0 \times 10^{4}$ sec
and $X_{c}$ being varied as a parameter ($0.5 \leq X_{c} \leq 1.5$), where
$\tau_{0}^{*} = 5.27 \times 10^{-6}$ sec, $\zeta = 1.5 \times 10^{-4}$,
$\delta = 0.81$, $b/\delta = 0.198$, and $a_{eff} = 0.55$.  The dotted lines
denote the least-squares fitting curves of the $H$ vs $T$ relation for each
$X_{c}$ to Eq.(\ref{eq01}) with $p \approx 1.5$.  (b) $H$-$T$ line. 
The value of $H$ as a function of $T$ is calculated from Eq.(\ref{eq17})
with $X_{c} = 0.28$ and $t_{W} = 100$ sec, where the values of
$\tau_{0}^{*}$, $\zeta$, $\delta$, $b/\delta$, and
$a_{eff}$ are the same as those for (a).  The line $T_{g0}(H)$
(d$\Delta\chi$/d$T$ vs $T$) ({\Large $\bullet$}) of Fig.~\ref{fig02} is
also shown for comparison.}
\end{figure}

In summary , we find a scaling relation that all the set of $Y =
R_{cr}/R_{H}$ vs $X = R_{W}/R_{H}$ obtained under the $H$-shift protocol
fall well on a single scaling curve.  The curve of $Y$ vs $X$ clearly shows
a gradual dynamic crossover from the relation $Y = X$ for $X\ll 1$ to $Y
\approx 1$ for large $X$.  In this sense, the $H$-shift aging process is
nothing but a dynamic crossover from the SG state in $H = 0$ to the
paramagnetic state in $H>0$ in the equilibrium limit.

What is the relation between $H$ and $T$, depending on the value of
$X_{c}$?  Note that $X_{c}$ $\approx 0.1$ for the deviation of $Y$ from the
linear relation $Y = X$, and $X_{c} \approx 1.6$ for the saturation of $Y$. 
From the relation $X = X_{c}$ in Eq.(\ref{eq15a}), the magnetic field $H$
can be derived as
\begin{equation}
H \approx X_{c}^{1/\delta}\zeta^{1/\delta}(1-T/T_{c})^{a_{eff}}
(t_{W}/\tau_{0}^{*})^{-(b/\delta)T/T_{c}} .
\label{eq17}
\end{equation}
In Fig.~\ref{fig12}(a) we show the plot of $H$ vs $T$ defined by
Eq.(\ref{eq17}) as $X_{c}$ is varied as a parameter ($0.5 \leq X_{c} \leq
1.5$), where $t_{W} = 1.0 \times 10^{4}$ sec, $\tau_{0}^{*} = 5.27 \times
10^{-6}$ sec, $\zeta = 1.5 \times 10^{-4}$, $\delta = 0.81$, $b/\delta =
0.198$, and $a_{eff} = 0.55$.  It seems that the $H$-$T$ diagram for the
fixed $X_{c}$ is similar to Eq.(\ref{eq01}) with $p = 3/2$ (the AT
prediction).  In fact, the least squares fit of these data of $H$ vs $T$ to
Eq.(\ref{eq01}) with $t_{W} = 1.0 \times 10^{4}$ sec for $2 \leq T \leq
3.6$ K yields the parameters $A$ and $p$.  The value of $p$ is equal to
$1.50 \pm 0.03$, independent of $X_{c}$, while the value of $A$ increases
as increasing $X_{c}$: $A = (8.20 \pm 0.15)X_{c}^{1.31 \pm 0.06}$ [kOe].

Finally we discuss the nature of the line $T_{g0}(H)$ (d$\Delta\chi$/d$T$
vs $T$) shown in Fig.~\ref{fig02}.  This line is well described by
Eq.(\ref{eq01}) with $p = 1.52 \pm 0.10$ and $A = 3.38 \pm 0.67$ kOe.  If
the above model is applicable to this line, the value of $X_{c}$ is
estimated as $X_{c} = 0.50$.  Figure \ref{fig12}(b) shows the plot of $H$
vs $T$, where $H$ is calculated from Eq.(\ref{eq17}) for each $T$.  Here we
choose $t_{W} = 100$ sec which is appropriate for the measurements of
$\chi_{ZFC}$ and $\chi_{FC}$, and $X_{c} = 0.28$ which is rather smaller
than $X_{c} = 0.50$.  We find that the line $T_{g0}(H)$ (d$\Delta\chi$/d$T$
vs $T$) for $2.5 < T < 3.6$ K lie well on the prediction from
Eq.(\ref{eq17}).  This result also suggests that the line $T_{g0}(H)$
(d$\Delta\chi$/d$T$ vs $T$) is also explained in terms of the droplet
picture.

\section{CONCLUSION}
Cu$_{0.5}$Co$_{0.5}$Cl$_{2}$-FeCl$_{3}$ graphite intercalation compound is
a three-dimensional short-range Ising SG with a spin freezing temperature
$T_{c}$ ($= 3.92 \pm 0.11$ K).  The stability of the SG phase in the
presence of a magnetic field $H$ has been studied from the spin freezing
temperature $T_{f}(\omega,H)$ and the peak time $t_{cr}(T,H)$ of $S(t)$ vs
$t$ for $T<T_{c}$.  We find that both $T_{f}(\omega,H)$ and $t_{cr}(T,H)$
show certain scaling behavior predicted from the droplet picture.  The
ratio $Y$ ($= R_{cr}/R_{H}$) is well described by a scaling function of $X$
($= R_{W}/R_{H}$) in the form of $Y=X-cX^{1+1/\delta}$ with $c = 0.71 \pm
0.06$ and $\delta = 0.81 \pm 0.07$.  This result indicates the instability
of the SG phase in the thermal equilibrium in a finite magnetic field: a
dynamic crossover occurs from SG dynamics to paramagnetic behavior.

\begin{acknowledgments}
We would like to thank H. Suematsu for providing us with single 
crystal kish graphite, and T. Shima and B. Olson for their assistance 
in sample preparation and x-ray characterization.
\end{acknowledgments}

\end{document}